\documentclass[aps,prl,twocolumn,groupedaddress]{revtex4-1}
\usepackage{graphicx}

\begin{document}


\title{The analytic radial acceleration relation for galaxy clusters}


\author{Man Ho Chan, Ka Chung Law}
\affiliation{Department of Science and Environmental Studies, The Education University of Hong Kong \\ 
Tai Po, New Territories, Hong Kong, China}


\date{\today}

\begin{abstract}
Recently, a tight correlation between the dynamical radial acceleration and the baryonic radial acceleration in galaxies - the radial acceleration relation - has been discovered. This has been claimed as an indirect support of the modified gravity theories. However, whether the radial acceleration relation could also be found in galaxy clusters is controversial. In this article, we derive and present an analytic radial acceleration relation for the central region of galaxy clusters. We examine the data of some large galaxy clusters and we find that the resulting radial acceleration relation has a very large scatter. Moreover, although the radial acceleration relation for galaxy clusters shows some agreement with the one discovered in galaxies for a certain range of baryonic radial acceleration, their functional forms are somewhat different from each other. This suggests that the radial acceleration relation may not be a universal relation in general.
\end{abstract}

\maketitle


\section{Introduction}
In the past few decades, many close relations between dark matter and baryonic matter were discovered, such as the Tully-Fisher relation \cite{Tully}, the Faber-Jackson relation \cite{Faber}, the mass-discrepancy acceleration relation \cite{McGaugh}, and the radial acceleration relation (RAR) \cite{McGaugh2,Lelli}. The close relations between dark matter and baryons are not intuitively expected because the standard cold dark matter model suggests that the interaction between dark matter and baryons (except gravity) is almost negligible. Therefore, although some studies are able to reproduce these relations using the cold dark matter framework \cite{Chan2,Ludlow,Stone}, the close relations between dark matter and baryons generally more favor the modified gravity theories because many of these theories suggest that baryonic mass and dynamical mass are inter-related \cite{Milgrom,Li,Green,Islam,Brouwer}.

Nevertheless, if any modified gravity theory is true for accounting the close relations between dark matter and baryons, those close relations should also be found in other structures such as galaxy clusters. It is because the modified gravity theory should be universal in nature so that the resulting relations between dark matter and baryons must be universal as well. For example, some studies have shown that the Faber-Jackson relation can also be found in galaxy clusters \cite{Sanders,Tian2}, which may give an indirect support of the modified gravity theories. In particular, many studies are now focusing on the RAR in galaxies because it reveals the existence of an apparent universal acceleration scale, which has been predicted by some modified gravity theories such as the Modified Newtonian Dynamics (MOND) \cite{Milgrom2} and the Emergent Gravity \cite{Verlinde}. However, whether a tight RAR could also be found in galaxy clusters is quite controversial. Some studies show that the functional form of the RAR in galaxies might still be true for galaxy clusters, although the acceleration scale value is different from that in galaxies \cite{Edmonds,Tian}. Some other studies show that the scatter of the RAR in galaxy clusters is too large to be consistent with that in galaxies \cite{Chan,Pradyumna,Gopika}. 

In this article, we derive an analytic RAR for the central region of galaxy clusters so that we can easily analyze the dependence of the RAR in galaxy clusters. We will also examine the analytic RAR by using the data of some large galaxy clusters, which can explicitly reveal the alleged acceleration scale and the scatter of the RAR in galaxy clusters.

\section{The analytic radial acceleration relation}
The dynamical radial acceleration is defined as
\begin{equation}
a_{\rm dyn}=\frac{GM_{\rm dyn}}{r^2},
\end{equation}
where $M_{\rm dyn}$ is the enclosed dynamical mass at radius $r$. Although the hot gas in a galaxy cluster is a pressure-supported system, we can define the baryonic radial acceleration by
\begin{equation}
a_{\rm bar}=\frac{GM_{\rm bar}}{r^2},
\end{equation}
where $M_{\rm bar}$ is the enclosed baryonic mass at radius $r$. This can be interpreted as the radial acceleration contributed by the baryonic mass if the gas pressure does not exist. Generally speaking, for large and massive galaxy clusters, the enclosed baryonic mass can be approximately represented by the enclosed hot gas mass $M_{\rm gas}$. However, for small galaxy clusters, the effect of stellar mass $M_{\rm star}$ could be significant, especially in their central regions \cite{Chiu}. Therefore, we include stellar mass in our calculation of the baryonic mass: $M_{\rm bar}=M_{\rm gas}+M_{\rm star}$.

\subsection{Hydrostatic dynamical mass framework}
When the hot gas with temperature $T$ is in hydrostatic equilibrium, we have
\begin{equation}
\frac{d}{dr}[n(r)kT]=-\frac{GM_{\rm dyn}(r)\rho(r)}{r^2},
\end{equation}
where $n(r)$ and $\rho(r)$ are the number density profile and mass density profile of the hot gas respectively. Generally speaking, the temperature of the hot gas is almost constant for most non-cool-core clusters \cite{Vikhlinin,Reiprich2}. The temperature gradient in the hot gas can be less than 8\% \cite{Hudson}. Therefore, it would be a very good approximation if we simply take a constant average temperature $T$ \cite{Chen}. Based on this assumption, Eq.~(3) can be rewritten as \cite{Chen}
\begin{equation}
M_{\rm dyn}(r)=-\frac{kTr}{G\mu m_p} \frac{d \ln n(r)}{d \ln r},
\end{equation}
where $\mu=0.59$ is the molecular weight and $m_p$ is the proton mass. 

On the other hand, observations indicate that the hot gas number density profile can be well described by the $\beta$ model \cite{Cavaliere,Reiprich,Chen}:
\begin{equation}
n(r)=n_0 \left(1+ \frac{r^2}{r_c^2} \right)^{-3\beta/2},
\end{equation}
where $n_0$ is the central number density, $r_c$ is the hot gas core radius and $\beta$ is the index parameter. Therefore, combining Eq.~(4) and Eq.~(5), and following Eq.~(1), we get the dynamical radial acceleration expression
\begin{equation}
a_{\rm dyn}(r)=\frac{3 \beta kTr}{\mu m_p(r^2+r_c^2)}.
\end{equation}
Generally speaking, there are two possible positions $r$ which can have the same value of $a_{\rm dyn}$ (see the example of the Coma cluster in Fig.~1):
\begin{equation}
r=\frac{3\beta kT \pm \sqrt{9\beta^2k^2T^2-4a_{\rm dyn}^2\mu^2m_p^2r_c^2}}{2\mu m_pa_{\rm dyn}}.
\end{equation}
In the followings, we only focus on the region for the smaller root of $r$ as it can be shown that the analytic relation derived would be converged in the smaller-root region only. This corresponds to the central region ($r \le r_c$) of the hot gas. As shown in Fig.~1, the values of $a_{\rm dyn}$ in the smaller-root region of a galaxy cluster can represent most of the possible values of the dynamical radial acceleration.  

Next, to simplify our discussion, we define a dimensionless term $y \equiv a_{\rm dyn}/a_{\rm max}$ with $a_{\rm max} \equiv 3\beta kT/2\mu m_pr_c$ so that the smaller root $r$ can be rewritten as
\begin{equation}
r=r_c\left(\frac{1}{y}-\sqrt{\frac{1}{y^2}-1} \right).
\end{equation}
For our focussing region $r \le r_c$, we have $y \le 1$. 

To get the hot gas mass profile, we can integrate the hot gas number density profile:
\begin{equation}
M_{\rm gas}(r)=\int_0^r 4 \pi r'^2m_gn(r')dr',
\end{equation}
where $m_g$ is the average mass of a hot gas particle. By substituting the number density profile in Eq.~(5), the above integral can be expressed analytically in terms of the generalized hypergeometric series. We write the generalized hypergeometric series explicitly by the sum of the following infinite series: 
\begin{equation}
M_{\rm gas}(r)=4 \pi m_gn_0 \frac{r^3}{3} \sum_{j=0}^{\infty} \frac{(3\beta/2)_j(3/2)_j}{(5/2)_jj!} \left(-\frac{r^2}{r_c^2} \right)^j,
\end{equation}
where $(x)_j=\Gamma(x+j)/\Gamma(x)=x(x+1)...(x+j-1)$ is the Pochhammer symbol. Since $(3/2)_j/(5/2)_j=3/(3+2j)$, following the definition of $a_{\rm gas}=GM_{\rm gas}/r^2$, we get:
\begin{equation}
a_{\rm gas}=4 \pi Gm_gn_0r \sum_{j=0}^{\infty} \frac{(3\beta/2)_j}{j!(3+2j)} \left(-\frac{r^2}{r_c^2} \right)^j.
\end{equation}
 
Substituting the expression of the smaller root $r$ from Eq.~(8) into Eq.~(11) and expand the series about $y=0$, we get the analytic series of $a_{\rm gas}$ up to the fifth-order terms:
\begin{eqnarray}
a_{\rm gas}&=&4 \pi Gm_gn_0r_c \left[\frac{1}{6}y+\left(\frac{1}{24}-\frac{1}{80} \beta \right)y^3 \right.\nonumber\\ 
&&\left.+\left(\frac{1}{48}-\frac{111}{4480} \beta+ \frac{9}{1792} \beta^2 \right)y^5+O(y^7) \right].  
\end{eqnarray}

Since $y \equiv a_{\rm dyn}/a_{\rm max}$, we can see that $a_{\rm dyn}$ and $a_{\rm gas}$ are explicitly related by Eq.~(12). For large and massive galaxy clusters, since $M_{\rm bar} \approx M_{\rm gas}$, Eq.~(12) can be regarded as the analytic RAR for the central region of galaxy clusters. Nevertheless, if the contribution of stellar mass in galaxy clusters is significant, then we need to revise Eq.~(12) to include the stellar mass component. Previous studies have shown that the stellar mass in galaxy clusters can be best described by a power law of hot gas mass \cite{Giodini}. Here, since we have focused on the central region of galaxy clusters, we use the data of the inner region obtained in \cite{Lagana} to examine the power-law relation between $M_{\rm bar}=M_{\rm star}+M_{\rm gas}$ and $M_{\rm gas}$. We find that $M_{\rm bar}$ can be best described by the following power law: $M_{\rm bar}/10^{13}M_{\odot}=\alpha_0(M_{\rm gas}/10^{13}M_{\odot})^{\alpha}$ with $\alpha_0=1.35^{+0.02}_{-0.03}$ and $\alpha=0.79 \pm 0.02$ (see Fig.~2). Putting it into Eq.~(12), we get:
\begin{equation}
a_{\rm bar}=a_b \left[4 \pi \sum_{j=0}^{\infty} \frac{(-1)^j(3\beta/2)_j}{j!(3+2j)} \left(\frac{1-\sqrt{1-y^2}}{y} \right)^{2j+3-2/\alpha} \right]^{\alpha},
\end{equation}
where $a_b=\alpha_0G(10^{13}M_{\odot})^{1-\alpha}(m_gn_0r_c^3)^{\alpha}/r_c^2$. We can expand Eq.~(13) around $y=0$ up to less than the fifth order, and substitute $\alpha_0 \approx 1.35$ and $\alpha \approx 0.79$ to get the following approximated form:
\begin{eqnarray}
a_{\rm bar}&\approx &a_b \left[2.4y^{0.37}+(0.22-0.14\beta)y^{2.37} \right.\nonumber\\ 
&&\left.+\left(0.093-0.26\beta +0.056 \beta^2 \right)y^{4.37} \right].  
\end{eqnarray}

Simply looking at Eq.~(14), it seems that $a_{\rm dyn}$ is tightly correlated with $a_{\rm bar}$, while the actual RAR depends on the empirical hot gas parameters (e.g. $r_c$, $T$, $\beta$ and $n_0$) of different galaxy clusters.

\subsection{NFW framework}
The analytic relation in Eq.~(14) is based on the hydrostatic equilibrium of hot gas. However, numerical simulations show that the density profile for cold dark matter can be best described by the Navarro-Frenk-White (NFW) profile \cite{Navarro}:
\begin{equation}
\rho_{\rm DM}=\frac{\rho_sr_s^3}{r(r+r_s)^2},
\end{equation}
where $\rho_s$ and $r_s$ are the scale density and scale radius of dark matter respectively. The enclosed dark matter mass profile can be obtained analytically: 
\begin{equation}
M_{\rm DM}=\int_0^r 4\pi r'^2\rho_{\rm DM}dr'=4\pi \rho_sr_s^3 \left[\ln \left(1+\frac{r}{r_s} \right)-\frac{r}{r+r_s} \right].
\end{equation}
Some studies have shown that the NFW profile can give good fit to the dynamical mass of galaxy clusters \cite{Pointecouteau}. Nevertheless, the enclosed dynamical mass profile ($M_{\rm DM}+M_{\rm bar}$) derived in the NFW framework is somewhat different from the hydrostatic dynamical mass in Eq.~(6). Therefore, in the followings, we also derive an analytic RAR relation within the NFW framework.

The dynamical radial acceleration is given by $a_{\rm dyn}=a_{\rm bar}+a_{\rm DM}$, where $a_{\rm DM}=GM_{\rm DM}/r^2$. Note that the expressions of $a_{\rm bar}$ and $a_{\rm gas}$ in the NFW framework are identical to those in the hydrostatic framework because the hot gas number density profile in Eq.~(5) is merely empirical. We first expand $a_{\rm DM}$ in terms of an infinite series:
\begin{equation}
a_{\rm DM}=4 \pi G \rho_sr_s \left[ \sum_{j=1}^{\infty} \left(1- \frac{1}{j} \right) \left(-\frac{r}{r_s} \right)^{j-2} \right].
\end{equation}
As we will focus on the central region, we expand the expression of $a_{\rm gas}$ in Eq.~(11) and approximate it by neglecting the fifth-order or higher terms:
\begin{equation}
a_{\rm gas} \approx 4\pi Gm_gn_0 \left(\frac{r}{3}-\frac{3\beta}{10r_c^2}r^3 \right).
\end{equation}
By using the relation $a_{\rm bar}=\alpha_0(10^{13}M_{\odot}G/r^2)^{1-\alpha}a_{\rm gas}^{\alpha}$, we can write
\begin{equation}
9\beta \left(\frac{r}{r_c} \right)^{5-2/\alpha}-10\left(\frac{r}{r_c} \right)^{3-2/\alpha}+\frac{15}{2\beta \pi} \left(\frac{a_{\rm bar}}{a_b} \right)^{1/\alpha}=0.
\end{equation}

We can go analytically further by putting $\alpha \approx 0.8=4/5$ into Eq.~(19). Expanding Eq.~(19) around $r/r_c=1$ up to the second order and solve the quadratic equation in $(r/r_c-1)$, we get
\begin{equation}
\frac{r}{r_c}=1-2\gamma[1+f(a_{\rm bar})]
\end{equation}
with 
\begin{equation}
f(a_{\rm bar})=\sqrt{1+\frac{4}{5\gamma}\left[\frac{8(1-4\gamma)}{1-3\gamma}-\frac{5(1-3\gamma)}{3(1+\gamma) \pi}\left(\frac{a_{\rm bar}}{a_b} \right)^{5/4} \right]},
\end{equation}
where $\gamma=(9\beta-2)/(27\beta+2)$. Putting Eq.~(20) into Eq.~(17) and following our definition $a_{\rm dyn}=a_{\rm DM}+a_{\rm bar}$, we have
\begin{eqnarray}
a_{\rm dyn}&=&a_{\rm bar}+4\pi G\rho_sr_s \nonumber\\ 
&& \times \sum_{j=1}^{\infty} \left(1-\frac{1}{j} \right) \left\{-\frac{r_c}{r_s}\left[1-2\gamma(1+f(a_{\rm bar}))\right] \right\}^{j-2}.  
\end{eqnarray}
Since we will focus on the central region $r<r_c$, we expand the first 3 terms ($j=1,2,3$) to get the final approximate analytic RAR $a_{\rm dyn}(a_{\rm bar})$ for the NFW framework:
\begin{equation}
a_{\rm dyn} \approx a_{\rm bar}+2\pi G\rho_sr_s \left[1-\frac{4r_c}{3r_s}+\frac{8r_c}{3r_s} \gamma (1+f(a_{\rm bar})) \right].
\end{equation}

\section{Data analysis}
To examine the analytic radial acceleration relation explicitly, we plot the RAR by putting the values of parameters from the available data. We analyze the data from the extended HIFLUGCS galaxy cluster sample obtained from \cite{Chen}. However, since we have assumed a constant temperature profile in the derivation, we will only use the data of the non-cool-core clusters. Also, we neglect the small galaxy clusters with $r_c \le 100$ kpc because the central bright cluster galaxies (BCGs) would dominate the baryonic mass at the centers, which might have complex behaviors. Therefore, we altogether analyze 52 non-cool-core clusters with core radii $r_c>100$ kpc in the extended HIFLUGCS sample. 

Besides, the fitted parameters from observations may not be accurate in describing the deep central region (e.g. $r \le 20$ kpc). Therefore, for the hydrostatic framework, we constrain our analysis to the range of $y=0.4-0.9$ (i.e. $r \approx 0.21r_c-0.63r_c$) only. For $y=0.9$, the error of our analytic RAR is less than 10\%. For the NFW framework, as we have expanded our series around $r=r_c$, we will focus on the region $r=0.5r_c-1r_c$. 

We first plot the RAR for the hydrostatic framework by using the data of 52 galaxy clusters (with gas mass $\sim 10^{13}M_{\odot}-10^{15}M_{\odot}$) with their corresponding parameters $\beta$, $r_c$, $T$ and $n_0$ (see Fig.~3). The values of the parameters can be found in \cite{Chen}. We have rescaled the values of the parameters by adopting the Hubble parameter $h=0.68$. The error bars of our results shown in Fig.~3 represent the possible ranges of $a_{\rm dyn}$ and $a_{\rm bar}$ due to the uncertainties of the input parameters. For $y=0.4-0.9$, we can see that the resultant RAR scatters in a very large $\log a_{\rm dyn}-\log a_{\rm bar}$ space. Although it is in remarkable agreement with the galactic RAR for a certain range of $a_{\rm bar}$, its functional form $a_{\rm dyn}(a_{\rm bar})$ is somewhat different from that of the galactic RAR (the galactic RAR has a smaller slope for larger $a_{\rm bar}$). The difference becomes more significant for a larger value of $y$ (e.g. $y=0.9$, see Fig.~3). Also, our results give a larger value of the `acceleration scale' (if it exists) for the central RAR in galaxy clusters compared with the galactic RAR. These results are consistent with other recent studies of the RAR in galaxy clusters \cite{Chan,Tian,Pradyumna,Edmonds}. 

For the NFW framework, we can calculate the NFW scale density $\rho_s$ and the scale radius $r_s$ from the X-ray parameters based on the analytic formulas recently obtained in \cite{Chan3}. The RAR calculated in the NFW framework generally has larger $a_{\rm dyn}$ compared with the one in the hydrostatic framework. This is because the NFW profile is a cusp profile, which predicts a higher central dark matter density and a larger dynamical acceleration. For the hydrostatic framework, the central dynamical profile is close to a constant cored profile, which predicts a smaller dynamical acceleration. Generally speaking, these two models provide two extreme descriptions of the central RAR profile. The actual distribution of the RAR would lie between these two benchmark models. 

We also plot the RARs of two particularly chosen galaxy clusters (the Coma cluster and the A2877 cluster, see Fig.~4). We can see that the two RARs have completely different functional forms (i.e. different slopes in the $\log a_{\rm dyn}-\log a_{\rm bar}$ space) compared with the existing RAR. The RAR in the NFW framework is almost `flat' because the dark matter contribution has dominated the dynamical mass. Since the NFW mass profile goes like $r^2$ in the central region, we get $a_{\rm dyn} \approx a_{\rm DM}=GM_{\rm DM}/r^2 \approx$ constant.  

\section{Discussion}
In this article, we have derived two analytic RARs (one for the hydrostatic framework and one for the NFW framework), which are applicable for the central region of galaxy clusters. The analytic RARs are particularly good for describing large and massive non-cool-core clusters. In fact, many previous related studies focus on the outer region of galaxy clusters (e.g. $r \ge r_c$) \cite{Tian,Chan}. Therefore, our study can give some new insight for understanding the behavior of the RAR for galaxy clusters, especially for their central region. Also, the derived analytic RAR can explicitly reveal the potential dependence of the RAR, and how the hot gas parameters affect the scatter and the functional form of the RAR. 

Although the functional form of the analytic RAR is quite simple, it depends on four empirical hot gas parameters $\beta$, $r_c$, $T$ and $n_0$, which have broad ranges of values in general. Therefore, it is not surprising that the resultant RAR for the central region would scatter in a large space. In the derivation of the hydrostatic framework, we have assumed that the hot gas temperature is constant and the hot gas is in hydrostatic equilibrium. The former assumption is very good for the non-cool-core clusters as they do not have any large temperature gradient near the central region \cite{Hudson}. For the latter one, hydrostatic equilibrium is also a very good assumption for the non-cool-core clusters because the temperature gradient is so small such that no convection would be expected to affect the hydrostatic equilibrium \cite{Reiprich2}. Another recent study shows that the systematic uncertainty for assuming hydrostatic equilibrium is less than 15\% in general \cite{Biffi}, which is relatively small compared with the possible scatter of the RAR. Besides, we have also examined the analytic RAR for the NFW framework. Therefore, our derived analytic expressions would be comprehensive enough in describing the RAR for large non-cool-core clusters. This also provides a comparison of the RARs between two different frameworks.

Generally speaking, although our derived RAR is only good for large non-cool-core clusters and we only focus on the central region, the resultant RAR scatters in a very large $\log a_{\rm dyn}-\log a_{\rm bar}$ space, although it shows some agreement with the galactic RAR for a certain range of $a_{\rm bar}$. Moreover, the functional form of the RAR for galaxy clusters is somewhat different from that of the galactic RAR, especially for larger values of $y$. The significant difference in functional form can also be seen from the plot using the data of individual galaxy clusters (e.g. the Coma cluster and the A2877 cluster in Fig.~4). These suggest that there may be no universal RAR for both galaxies and galaxy clusters. Besides, another recent study also suggests that there is a significant difference between the RARs of early-type and late-type galaxies \cite{Brouwer}. Therefore, the tight RAR shown in previous studies may be just a special characteristic for rotating system only. Nevertheless, since the uncertainties of the input parameters are quite large for galaxy clusters, more precise observational data are required to verify our conclusion. Note that our results do not have any implication for the modified gravity theories. However, if there is no universal RAR, the RAR discovered in galaxies may not be a good evidence for supporting any modified gravity theories which predict a universal acceleration scale (e.g. MOND).  

\begin{figure}
\vskip 4mm
 \includegraphics[width=80mm]{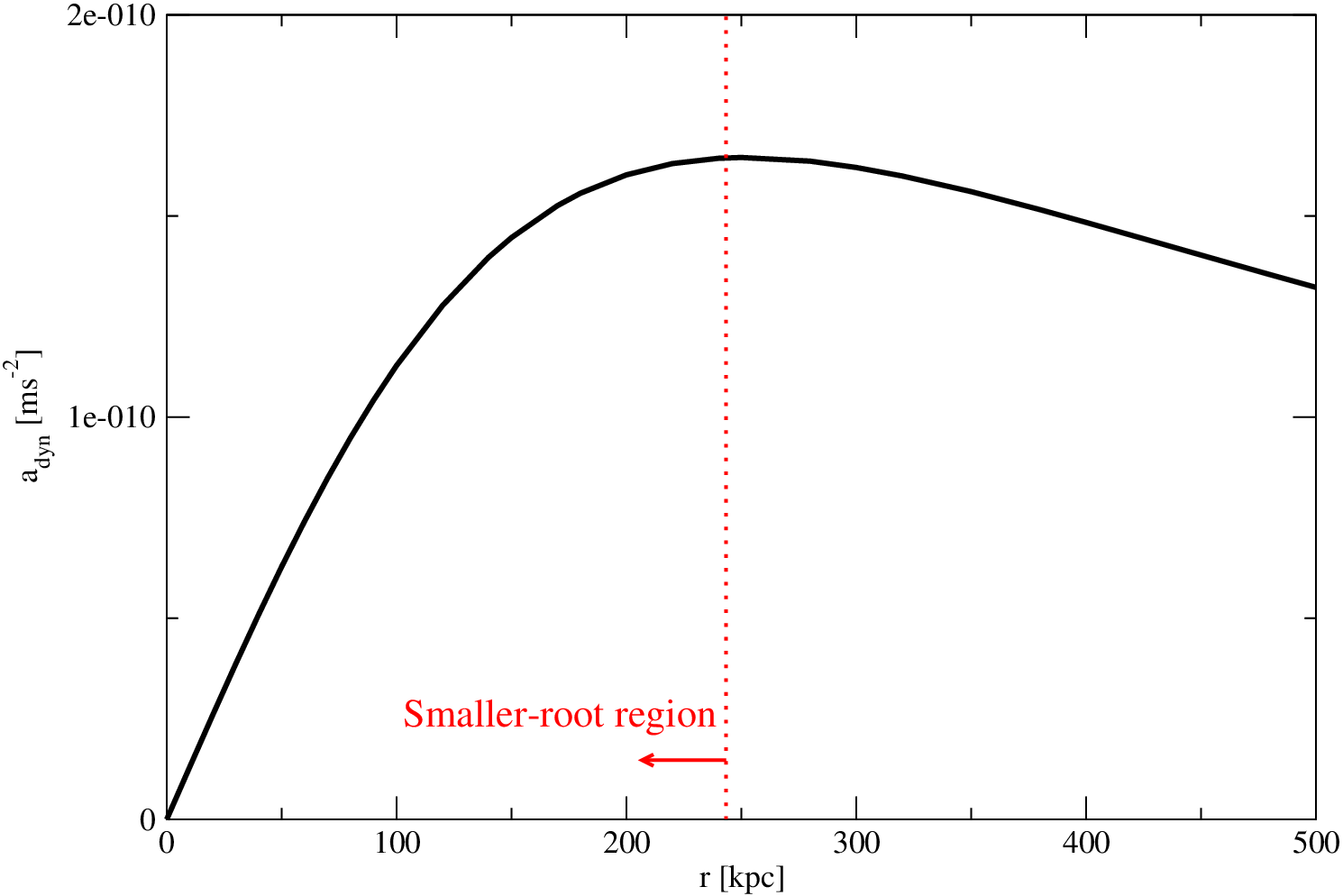}
 \caption{The black solid line represents the values of $a_{\rm dyn}$ against different radii $r$ from the center of the Coma cluster.}
\vskip 4mm
\end{figure}

\begin{figure}
\vskip 4mm
 \includegraphics[width=80mm]{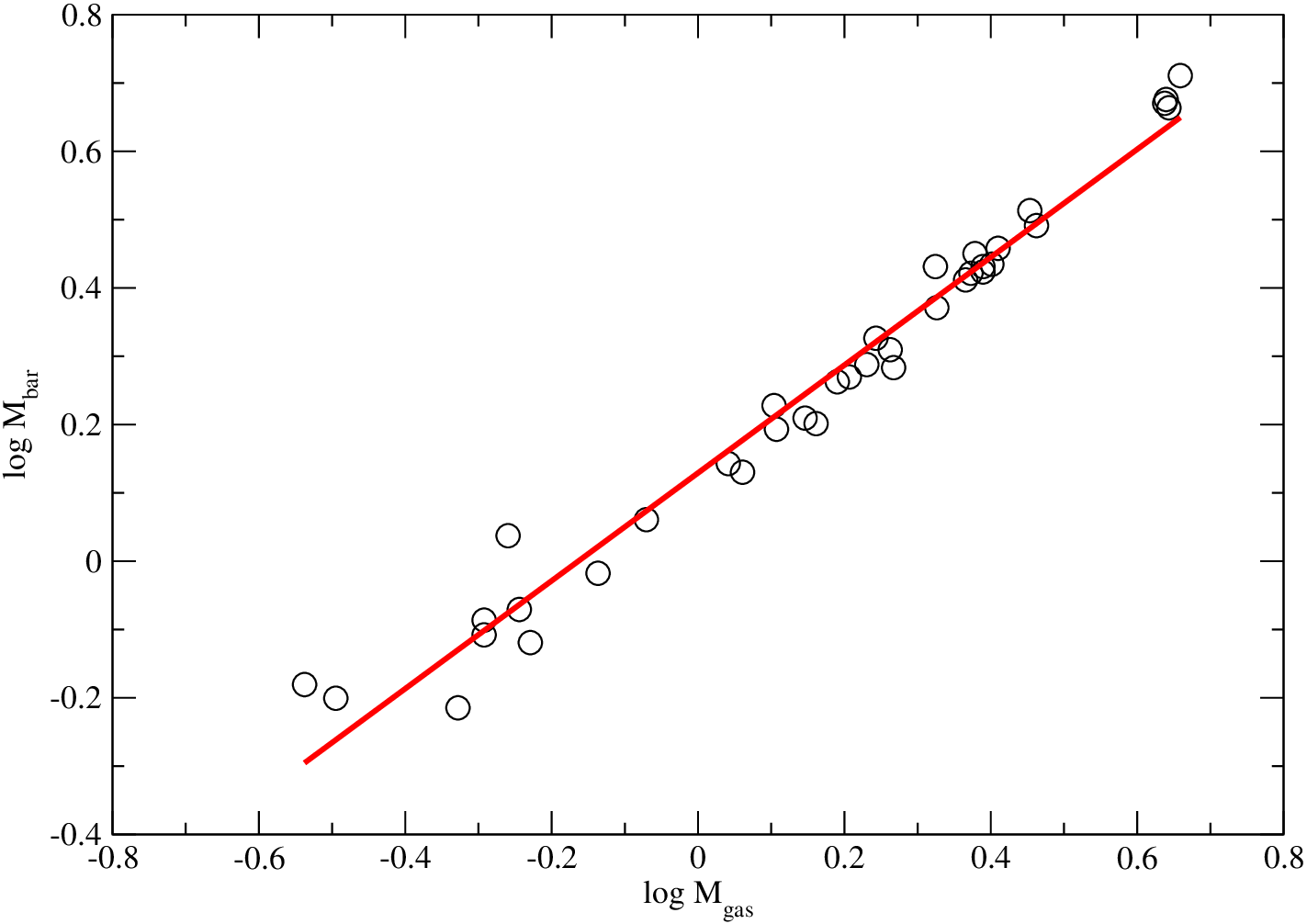}
 \caption{The circles are the data of $M_{\rm bar}=M_{\rm gas}+M_{\rm star}$ and $M_{\rm gas}$ at the inner region $r=r_{2500}$ of 37 galaxy groups and clusters obtained in \cite{Lagana} (in the unit of $10^{13}M_{\odot}$). The red line is the best-fit power-law relation $\log M_{\rm bar}=0.13+0.79\log M_{\rm gas}$.}
\vskip 4mm
\end{figure}

\begin{figure}
\vskip 4mm
 \includegraphics[width=80mm]{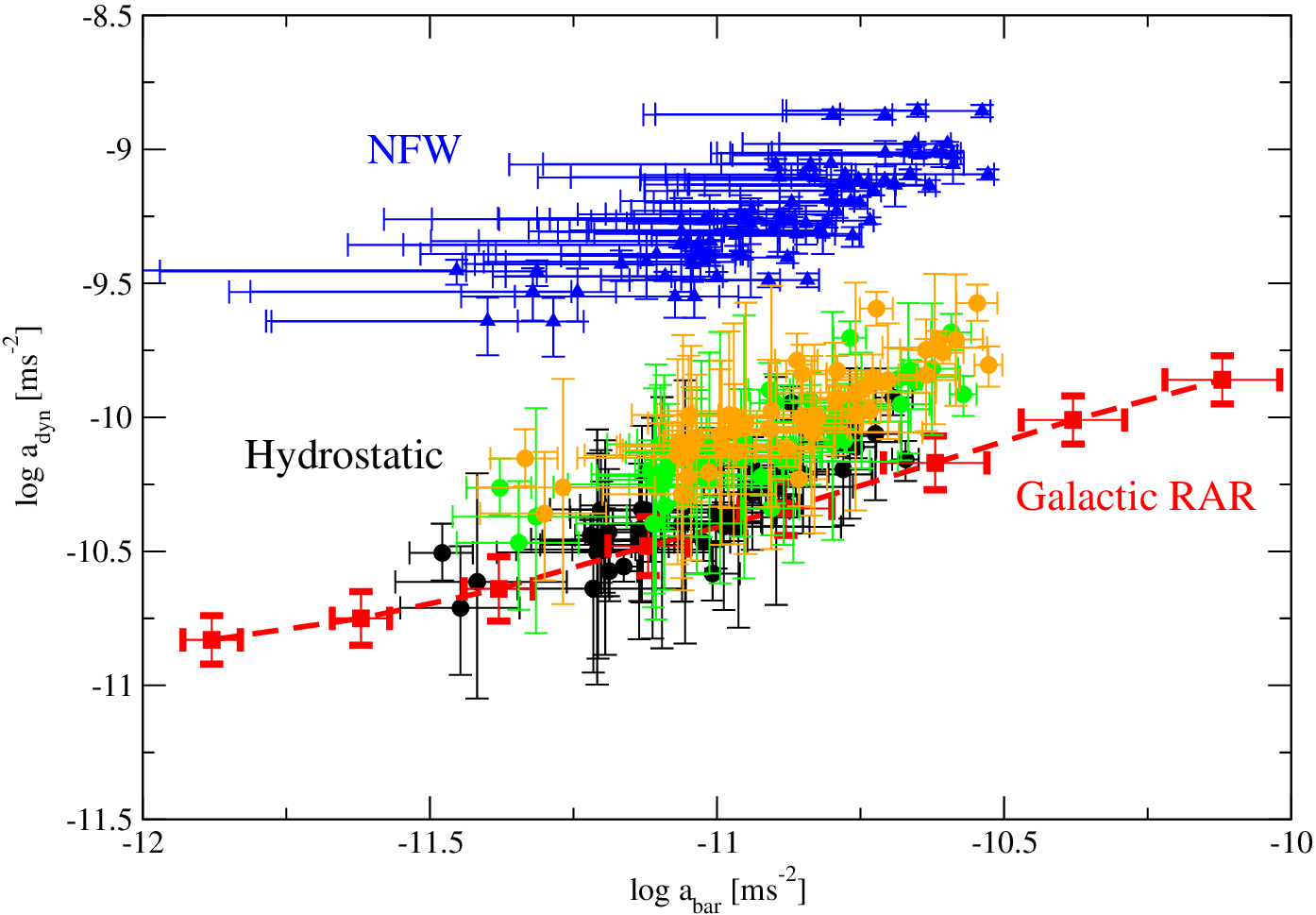}
 \caption{The circular dots represent the data of the RAR in the hydrostatic framework calculated by our derived analytic expression for $y=0.4$ (black), $0.7$ (green) and $0.9$ (orange). The blue triangles represent the data of the RAR in the NFW framework for $r=0.5r_c$ and $r=0.9r_c$. Here, the error bars associated with the circular dots and blue triangles represent the possible ranges of $a_{\rm dyn}$ and $a_{\rm bar}$ due to the uncertainties of the input parameters. The data of the 52 large non-cool-core galaxy clusters are derived from the X-ray parameters obtained in \cite{Chen}. The red squares with error bars linked up by the red dashed line represent the RAR of galaxies for comparison \cite{McGaugh2}.}
\vskip 4mm
\end{figure}

\begin{figure}
\vskip 4mm
 \includegraphics[width=80mm]{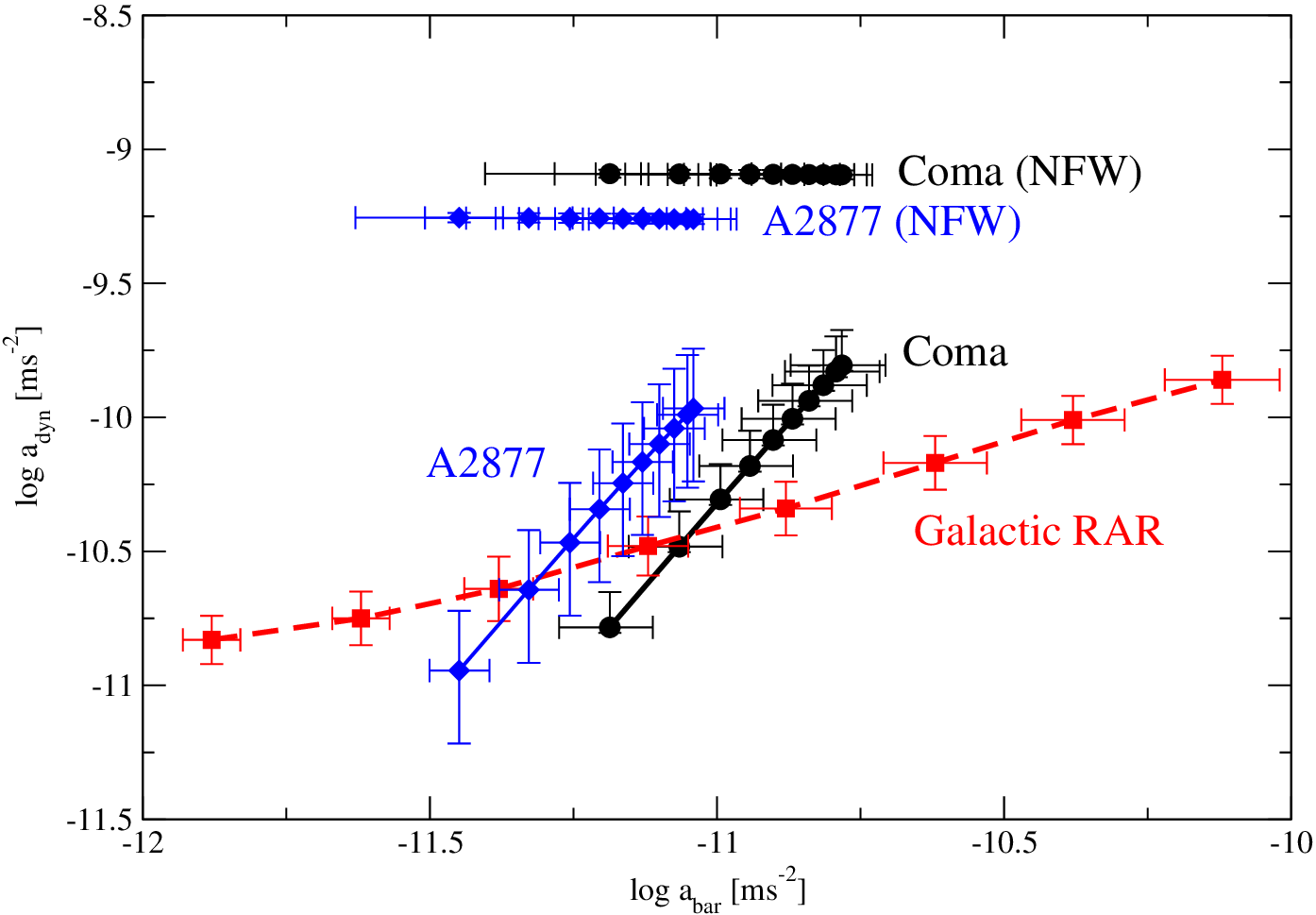}
 \caption{The black dots and blue diamonds represent the RARs for the Coma cluster and the A2877 cluster respectively. The red squares with error bars linked up by the red dashed line represent the RAR of galaxies for comparison \cite{McGaugh2}.}
\vskip 4mm
\end{figure}

\section{Acknowledgements}
We thank the anonymous referee for useful constructive feedbacks and comments. The work described in this paper was partially supported by the Seed Funding Grant (RG 68/2020-2021R) and the Dean's Research Fund of the Faculty of Liberal Arts and Social Sciences, The Education University of Hong Kong, Hong Kong Special Administrative Region, China (Project No.: FLASS/DRF 04628).


\begin{thebibliography}{00}
\bibitem{Tully} R. B. Tully \& J. R. Fisher, Astron. Astrophys. 54, 661 (1977).
\bibitem{Faber} S. M. Faber \& R. E. Jackson, Astrophys. J. 204, 668 (1976).
\bibitem{McGaugh} S. S. McGaugh, Astrophys. J. 609, 652 (2004).
\bibitem{McGaugh2} S. S. McGaugh, F. Lelli \& J. M. Schombert, Phys. Rev. Lett. 117, 201101 (2016).
\bibitem{Lelli} F. Lelli, S. S. McGaugh, J. M. Schombert \& M. S. Pawlowski, Astrophys. J. 836, 152 (2017).
\bibitem{Chan2} M. H. Chan, Int. J. Mod. Phys. D 26, 1750118 (2017).
\bibitem{Ludlow} A. D. Ludlow {\it et al.}, Phys. Rev. Lett. 118, 161103 (2017).
\bibitem{Stone} C. Stone \& S. Courteau, Astrophys. J. 882, 6 (2019).
\bibitem{Milgrom} M. Milgrom, Phys. Rev. Lett. 117, 141101 (2016).
\bibitem{Li} P. Li, F. Lelli, S. S. McGaugh \& J. Schombert, Astron. Astrophys. 615, A3 (2018).
\bibitem{Green} M. A. Green \& J. W. Moffat, Phys. Dark Uni. 25, 100323 (2019).
\bibitem{Islam} T. Islam \& K. Dutta, Phys. Rev. D 101, 084015 (2020).
\bibitem{Brouwer} M. M. Brouwer {\it et al.}, Astron. Astrophys. 650, A113 (2021).
\bibitem{Sanders} R. H. Sanders, Mon. Not. R. Astron. Soc. 407, 1128 (2010).
\bibitem{Tian2} Y. Tian, P.-C. Yu, P. Li, S. S. McGaugh \& C.-M. Ko, Astrophys. J. 910, 56 (2021).
\bibitem{Milgrom2} M. Milgrom, Astrophys. J. 270, 365 (1983).
\bibitem{Verlinde} E. P. Verlinde, SciPost Phys. 2, 016 (2017).
\bibitem{Edmonds} D. Edmonds, D. Minic \& T. Takeuchi, arXiv:2009.12915.
\bibitem{Tian} Y. Tian, K. Umetsu, C.-M. Ko, M. Donahue \& I.-N. Chiu, Astrophys. J. 896, 70 (2020).
\bibitem{Chan} M. H. Chan \& A. Del Popolo, Mon. Not. R. Astron. Soc. 492, 5865 (2020).
\bibitem{Pradyumna} S. Pradyumna, S. Gupta, S. Seeram \& S. Desai, Phys. Dark Uni. 31, 100765 (2021).
\bibitem{Gopika} K. Gopika \& S. Desai, Phys. Dark Uni, in press (arXiv:2106.07294).
\bibitem{Chiu} I. Chiu {\it et al.}, Mon. Not. R. Astron. Soc. 478, 3072 (2018).
\bibitem{Vikhlinin} A. Vikhlinin, A. Kravtsov, W. Forman, C. Jones, M. Markevitch, S. S. Murray \& L. Van Speybroeck, Astrophys. J. 640, 691 (2006).
\bibitem{Reiprich2} T. H. Reiprich, K. Basu, S. Ettori, H. Israel, L. Lovisari, S. Molendi, E. Pointecouteau \& M. Roncarelli, Sp. Sci. Rev. 177, 195 (2013).
\bibitem{Cavaliere} A. Cavaliere \& R. Fusco-Femiano, Astron. Astrophys. 49, 137 (1976).
\bibitem{Chen} Y. Chen, T. H. Reiprich, H. B\"ohringer, Y. Ikebe \& Y.-Y. Zhang, Astron. Astrophys. 466, 805 (2007).
\bibitem{Reiprich} T. H. Reiprich \& H. B\"ohringer, Astrophys. J. 567, 716 (2002).
\bibitem{Giodini} S. Giodini {\it et al.}, Astrophys. J. 703, 982 (2009).
\bibitem{Lagana} T. F. Lagana, N. Martinet, F. Durret, G. B. Lima Neto, B. Maughan \& Y.-Y. Zhang, Astron. Astrophys. 555, A66 (2013).
\bibitem{Navarro} J. F. Navarro, C. S. Frenk \& S. D. M. White, Astrophys. J. 490, 493 (1997).
\bibitem{Pointecouteau} E. Pointecouteau, M. Arnaud \& G. W. Pratt, Astron. Astrophys. 435, 1 (2005).
\bibitem{Chan3} M. H. Chan, Astrophys. J. 923, 95 (2021).
\bibitem{Hudson} D. S. Hudson, R. Mittal, T. H. Reiprich, P. E. J. Nulsen, H. Andernach \& C. L. Sarazin, Astron. Astrophys. 513, A37 (2010).
\bibitem{Biffi} V. Biffi {\it et al.}, Astrophys. J. 827, 112 (2016).
\end{thebibliography}
\end{document}